\documentclass{article}

\usepackage{amsmath} 
\usepackage{epsfig,graphics} 
\usepackage{color} 
\usepackage{amssymb} 
\usepackage{latexsym}

\newcommand{\MW}{\mathcal{W}}

\newcommand{\HD}{\hat{D}}

\newcommand{\la}{\langle} 
\newcommand{\ra}{\rangle}

\newcommand{\tO}{\textrm{otherwise}} 
\newcommand{\hD}{\hat{\Delta}} 
 
\newcommand{\Tr}{\textrm{Tr}}

\title{Non-positivity of Groenewold operators}
\author{A. J. Bracken and J. G. Wood \\
  Centre for Mathematical Physics, Department of Mathematics \\
  University of Queensland, Brisbane, Australia 4072.}

\begin{document}

\maketitle

\begin{abstract}
A central feature in the Hilbert space formulation of classical mechanics is
the quantisation of classical Liouville densities, leading to what may be
termed term Groenewold operators. We investigate the spectra of
the Groenewold operators that correspond to Gaussian and to certain uniform
Liouville densities. We show that when the classical coordinate-momentum
uncertainty product falls below Heisenberg's limit, the Groenewold operators in
the Gaussian case develop negative eigenvalues and eigenvalues larger than 1.
However, in the uniform case, negative eigenvalues are shown to persist for
arbitrarily large values of the classical uncertainty product.
\end{abstract}


\begin{section}{Introduction} 

The phase space formulation of quantum mechanics has its origins in the work 
of Weyl \cite{Weyl31} and Wigner \cite{Wigner32} more than 70 years ago. In
this  formulation, quantum observables are represented by phase space functions
and  quantum states by quasi-density (Wigner) functions that evolve in time
according to a  deformation of the classical Liouville equation. Although many
calculations are  more difficult to perform in the phase space context, this
formulation has the distinct advantage of the common language of phase space
for the purpose of comparing classical and quantum dynamics. Indeed, it was
for such a  purpose (the calculation of quantum corrections to thermodynamic
quantities)  that Wigner first introduced his now famous function. 

This begs the following question: if quantum mechanics can be described in the 
classical setting of phase space, can classical mechanics be described in the 
quantum setting of Hilbert space?  In fact, it was already recognised by
Groenewold \cite{Groenewold46}  nearly 60 years ago that this is possible and,
after related work in the interim \cite{Bopp61,Jordan61,Bayen78,Muga92,Sala94},
it has recently been made explicit  \cite{Bracken03}. The key to understanding
this is to recognise the unitarity, and hence the invertibility, of the
Weyl-Wigner transform \cite{Dubin00,Bracken03A}, which maps  quantum operators
into phase space functions. 

The inverse transform is Weyl's quantisation map, which transforms classical
functions (observables) into linear operators on Hilbert space.  Just as the
quantum density operator is mapped into a quasi-probability function on phase
space (a Wigner function) by the Weyl-Wigner transform,  so a classical
Liouville density function on phase space is mapped by the inverse transform
into a quasi-density operator on Hilbert space (which we suggest to call a
Groenewold operator).  Just as a Wigner function is normalised on all of phase
space, but is  not in general  a non-negative definite function, so a Groenewold
operator has unit trace, but is not in general a non-negative definite 
operator.  Just as quantum averages can be calculated in  the phase space
formulation of quantum mechanics using the Wigner function in the usual way of
a  density function, so classical averages can be calculated in the Hilbert
space formulation of classical mechanics using the Groenewold operator in the
usual way of a density operator.    Finally, just as the Wigner function
evolves in time in accordance with a deformation of classical mechanical
evolution, governed by a deformed Poisson bracket (the star  or Moyal bracket),
so the Groenewold operator evolves according to a deformation of quantum
mechanical evolution, governed by a deformed commutator bracket
\cite{Jordan61}, which has been called elsewhere the odot bracket
\cite{Bracken03}.  

The time evolution of Groenewold operators in Hilbert space can also be
described using a superoperator formalism that is particularly convenient for
numerical work, as shown by Muga {\it et. al.} \cite{Muga92, Sala94}.   More recently
\cite{Bracken03}, an expression has been derived for the classical evolution on
Hilbert space as a  series in increasing powers of $\hbar$ with the quantum
evolution as leading term. This form of the  evolution suggests the possibility
of probing the classical/quantum interface from  the quantum side, by
calculating classical corrections to a quantum  evolution, rather than quantum
corrections to a classical evolution.  

It is desirable to understand the spectra of Groenewold quasi-density operators
because of their primary role as the Hilbert space representatives of classical
states.  Some insight into how negative eigenvalues of Groenewold operators can
arise is provided by the recent observation [15] that if a nonlinear classical
evolution is applied to an initial quantum density operator corresponding to a
coherent state (such an operator is also a non-negative definite Groenewold
operator, corresponding to a Gaussian initial density on phase space), then the
evolved operator develops negative eigenvalues.  This is a special case of a
result obtained earlier [7] for initial density operators corresponding to more
general pure states. With regard to characterising those Groenewold operators
that are non-negative definite, we note that every non-negative Wigner function
can also be regarded as a Liouville density, and the corresponding true density
operator is therefore also a Groenewold operator that is non-negative definite. 
However, the problem of identifying all non-negative Wigner functions has been
solved only partially \cite{Hudson74}. The only pure states having non-negative
Wigner functions are  coherent states (modulo a linear canonical
transformation), and all such Wigner functions are Gaussians. Although progress
has been made in extending these results to mixed states
\cite{Narcowich88,Brocker95}, the extent of the set of all non-negative
mixed-state Wigner functions is still unknown. 

In what follows, we are concerned with properties of Groenewold  operators at a
fixed time, and not with their evolution.  We consider a classical system with
one linear degree of freedom,  in the phase plane $\Gamma$ and in Hilbert space
${\cal H}$.  We examine the eigenvalue spectra of Groenewold operators in
${\cal H}$  corresponding to (A) classical  densities that are Gaussian on
$\Gamma$ and (B)  classical  densities that are uniform on a circular or 
elliptical disk in $\Gamma$, and zero elsewhere.   In both cases the
eigenvalues can be calculated exactly. In Case (A), we show that  negative
eigenvalues, and eigenvalues that exceed 1, appear when the classical 
uncertainty product $\Delta q \Delta p$ falls below $\hbar/2$, the minimum
allowed by quantum mechanics, and we discuss the  limiting form  of the
spectrum as the Gaussian approaches a  delta-function.  

The reader might be misled into thinking that a general classical density will
give rise to a positive Groenewold operator whenever the uncertainty product is
greater than $\hbar/2$.   This is not so, and we show that in Case (B)  the
corresponding  Groenewold operators have negative eigenvalues for {\it all}
values of $\Delta q  \Delta p$. 

\end{section} 
 
\begin{section}{Groenewold quasi-density operators} 
 
We denote a general classical (Liouville) density on $\Gamma$ by $\rho(q,p)$.
It satisfies 
\begin{equation} 
\rho(q,p) \,\geq\,0\,, \quad \int_\Gamma \rho(q,p)\,=\,1\,. 
\label{pdo01} 
\end{equation} 
The corresponding Groenewold operator may be defined through the inverse 
Weyl-Wigner 
transform: 
\begin{equation} 
\hat{\rho} \,=\, 2\pi\hbar \MW^{-1}(\rho) \,=\, \int_\Gamma 
\rho(q,p)\hD(q,p) dq dp\,, 
\label{pdo02} 
\end{equation} 
where $\hD(q,p)$ is the Weyl-Wigner kernel \cite{Stratonovich57}, which may be 
defined as \cite{Royer77}
\begin{equation} 
\hD(q,p) \,=\, 2 \HD(q,p) \hat{P} \HD(q,p)^\dagger \,=\, 2\hD(2q,2p) \hat{P}\,. 
\label{pdo04} 
\end{equation} 
Here $\hat{P}$ denotes the parity operator and $\HD(q,p) = \exp(i(q\hat{p} -
p\hat{q})/\hbar)$ is  a unitary displacement operator. Note that the
eigenvalues of $\hD(q,p)$ are $\pm 2$ \cite{Cahill69} and since Groenewold
operators can be expressed as a convex combination of these operators as in
\ref{pdo02}, their eigenvalues lie in the range $[-2,2]$, whereas those for a
true density operator are restricted to the interval $[0,1]$.
 
If $A(q,p)$, $B(q,p)$ are classical functions on phase space and $\hat{A} = 
\MW^{-1}(A)$, $\hat{B}=\MW^{-1}(B)$ are their operator images under the inverse 
Weyl-Wigner transform, then we have  \cite{Hillery84} 
\begin{equation} 
\frac{1}{2\pi\hbar}\int_\Gamma A(q,p) dq dp \,=\, \Tr(\hat{A})\,, \quad 
\frac{1}{2\pi\hbar}\int_\Gamma A(q,p) B(q,p)dq dp \,=\, \Tr(\hat{A} 
\hat{B})\,. 
\end{equation} 
Note also that if $A(q,p)$ is real-valued, $\hat{A}$ is Hermitian. 
These properties ensure that a Groenewold operator $\hat{\rho} = 
\MW^{-1}(\rho)$ has a number of important features in common with true 
density operators: it is an Hermitian operator with unit trace \cite{Bracken03} 
and the trace of the product of two Groenewold operators satisfies 
\begin{equation} 
\Tr(\hat{\rho}\hat{\rho}') \,=\, 2\pi\hbar \int_\Gamma 
\rho(q,p)\rho'(q,p) dq dp \, \geq \, 0 \,. 
\label{pdo07} 
\end{equation} 
Furthermore, as mentioned above, the calculation of  classical averages on
Hilbert space takes the familiar form used for  quantum averages, with the
Groenewold operator acting in the role of a density operator:
\begin{equation} 
\la A \ra \,=\, \int_\Gamma \rho(q,p) A(q,p) dq dp\,=\, \Tr(\hat{\rho} 
\hat{A})\,. 
\label{pdo08} 
\end{equation} 
 
\end{section} 
 
\begin{section}{Gaussian densities} 
 
Despite the similarities between Groenewold operators and  true density
operators, there are important differences, the most obvious of which is the
failure of the Groenewold operator to be non-negative definite in general.

The uncertainty product $\Delta q\Delta p$ associated with a classical density
is non-negative but can be arbitrarily small, and it is not surprising to find
signatures of `non-quantum' behaviour in the Hilbert space representation of 
classical densities appearing when this uncertainty product is less than 
$\hbar/2$, the minimum value allowed in quantum mechanics.

In order to investigate this further, we introduce the class of Gaussian 
densities 
\begin{equation} 
\rho_{\beta,\gamma}^{(G)}(q,p) \,=\, \frac{1}{\pi \beta\gamma} 
e^{-(q^2/\beta^2+ p^2/\gamma^2)}\,, 
\label{gd01} 
\end{equation} 
where $\beta$, $\gamma$ are constants with units of  Length and Momentum
respectively.  The corresponding classical uncertainty is given by  $\Delta q
\Delta p = \beta\gamma /2$.  
 
From (\ref{pdo02}) we see that the corresponding Groenewold operators  are
given by 
\begin{equation} 
\hat{\rho}_{\beta,\gamma}^{(G)} \,=\, \int_\Gamma 
\rho_{\beta,\gamma}^{(G)}(q,p) \hD(q,p) d\Gamma\,. 
\label{gd02} 
\end{equation} 
We now introduce the complex-conjugate  variables $\alpha,\overline{\alpha}$
with $\alpha = (\sqrt{\gamma/\beta}\,q + i\sqrt{\beta/\gamma}\,p)/\sqrt{2\hbar}$
and the corresponding annihilation-creation operator pair
$\hat{a},\hat{a}^\dagger$ with $\hat{a}  =(\sqrt{\gamma/\beta}\,\hat{q} + 
i\sqrt{\beta/\gamma}\, \hat{p})/\sqrt{2\hbar}$, together  with the ``number
operator" $\hat{N}=\hat{a}^{\dagger}\hat{a}$. This allows us to define the Fock
basis 
\begin{equation} 
\{|n\ra, n=0,1,2,\ldots\}\,,\quad \hat{a} |0\ra \,=\, 0\,,\quad 
|n\ra \,=\, \frac{(\hat{a}^\dagger)^n}{\sqrt{n!}}|0\ra\,,\quad \hat{N}|n\ra 
\,=n|n\ra\,.   
\label{gd03} 
\end{equation} 
 
The reason for introducing this basis is that the Groenewold operators 
$\hat{\rho}_{\beta, \gamma}^{(G)}$ commute with $\hat{N}$ and are diagonal on the 
Fock states. This can be seen by a direct calculation of their matrix elements:
\begin{equation} 
\la n|\hat{\rho}_{\beta,\gamma}^{(G)} | m \ra \,=\, \overline{\la m
|\hat{\rho}_{\beta,\gamma}^{(G)} | n \ra} \,=\, \int_\Gamma  \rho_{\beta,
\gamma}^{(G)}(q,p) \la n |\hD(q,p)| m \ra dq dp\,, 
\label{gd04} 
\end{equation} 
where, in terms of the variables $\alpha,\overline{\alpha}$, $\la n | 
\hD(\alpha,\overline{\alpha})| m\ra$ is given by \cite{Cahill69} 
\begin{equation} 
\la n|\hD(\alpha,\overline{\alpha}) | m\ra \,=\, 2(-1)^m \sqrt{\frac{m!}{n!}} (2
\alpha)^{n-m}  L_m^{n-m}(4|\alpha|^2) e^{-2|\alpha|^2}\,, \quad n\geq m\,.
\label{gd05} 
\end{equation} 
 
If we introduce the polar-type variables $(t,\phi)$ with $\alpha = 
\sqrt{t/2}\exp(i\phi)$ and $\overline{\alpha} = 
\sqrt{t/2}\exp(-i\phi)$, then we may express (\ref{gd04}) as 
\begin{equation} 
\la n|\hat{\rho}_{\beta,\gamma}^{(G)} | m \ra \,=\, \frac{(-1)^m}{2 \pi
\beta\gamma}  \int_0^{2\pi} d\phi \,e^{-i(n-m)\phi}  \sqrt{\frac{m!}{n!}}
\int_0^\infty  dt\, (\sqrt{2}t)^{(n-m)/2} L_m^{n-m}(2t)
e^{-(1+\tfrac{\hbar}{\gamma\beta})t}\,.  \label{gd07} 
\end{equation} 
Except in the case that $n=m$, the integral over $\phi$ vanishes, and hence, 
$\hat{\rho}_{\beta, \gamma}^{(G)}$ is diagonal on the Fock states, with eigenvalues 
\begin{equation} 
\lambda_n^{(G)}(\beta,\gamma) \,=\, \frac{(-1)^n}{\beta \gamma} \int_0^\infty dt \,
e^{-(1+\tfrac{\hbar}{\beta \gamma})t} L_n(2t)\,,\quad n=0,\,1,\,2,\,\dots \,.
\label{gd08} 
\end{equation} 
Note that this result generalises in the sense that if $\hat{f} = \MW^{-1}(f)$, 
where $f(q,p) = f(q^2/\beta^2+ p^2/\gamma^2)$, then $[\hat{f},\hat{N}]=0$ and 
the eigenvalues of $\hat{f}$ satisfy a formula that generalises (\ref{gd08}). 
 
After evaluating (\ref{gd08}) \cite{Gradshteyn00}, we 
get the eigenvalues of $\hat{\rho}_{\beta,\gamma}^{(G)}$ in the 
simple form 
\begin{equation} 
\lambda_n^{(G)}(\beta, \gamma) \,=\, \frac{2\hbar}{\beta\gamma+\hbar} 
\left(\frac{\beta\gamma-\hbar}{\beta \gamma+\hbar}\right)^n\,,\quad n=0,\,1,\,2,\,\dots\,. 
\label{gd09} 
\end{equation} 
Since $\Delta q \Delta p=\beta\gamma/2$ and no quantum state exists in  which
$\Delta q \Delta p < \hbar/2$, one expects that non-quantum  features of
$\hat{\rho}_{\beta,\gamma}^{(G)}$ should be observed for $\beta\gamma<  \hbar$.
The result (\ref{gd09}) shows that this is  indeed the case, since for
$\beta\gamma <\hbar$ and  $n$ odd, the eigenvalue
$\lambda_n^{(G)}(\beta,\gamma) <0$, while at least one of the eigenvalues lies
above $1$. Hence for $\beta \gamma<  \hbar$, when the classical density
$\rho_{\beta,\gamma}^{(G)}(q,p)$  may be called ``strongly classical", the
Groenewold operator $\hat{\rho}_{\beta,\gamma}^{(G)}$ displays the  non-quantum
feature of negative eigenvalues and eigenvalues that exceed 1.  This is
analogous to the  negative  values exhibited by Wigner functions corresponding
to  ``strongly quantum states", such as excited Fock states \cite{Raymer97}. 
 
When the operator ${\hat \rho}^{(G)}_{\beta,\gamma}$  is considered on the
coordinate representation of Hilbert space, it appears as an integral operator
with kernel \cite{Bracken03A}
\begin{eqnarray}
\rho_{K\,\beta,\gamma}^{(G)}(x,y) &=&
\int_{-\infty}^{\infty}\rho_{\beta,\gamma}^{(G)}\left(
[x+y]/2,p\right)\,e^{ip(x-y)/\hbar}\,dp \nonumber\\
&=& \frac{1}{\beta\sqrt{\pi}}
\,e^{-(x+y)^2/(4\beta^2)}\,e^{-\gamma^2(x-y)^2/(4\hbar^2)}\,. \label{kerneldef}
\end{eqnarray}
Because the $\lambda_n^{(G)}(\beta,\gamma)$ are the eigenvalues of
${\hat\rho}^{(G)}_{\beta,\gamma}$,  it follows that 
\begin{equation}
\int_{-\infty}^{\infty}\rho_{K\,\beta,\gamma}^{(G)}(x,y)\varphi_n(y)\,dy
=\lambda_n^{(G)}(\beta,\gamma) \varphi_n(x)\,,
\label{identity1}
\end{equation}
where $\varphi_n(x)$ is the coordinate representative of the state $|n\rangle$.
This has the well-known form of an oscillator eigenfunction,
\begin{equation}
\varphi_n(x)= {\text const.} H_n( x\sqrt{\gamma/\beta\hbar}) \,e^{-\gamma
x^2/(2\beta\hbar)}\,, 
\label{hermite}
\end{equation}
where $H_n$ is the Hermite polynomial \cite{Abramowitz70}, and then
(\ref{kerneldef}) and (\ref{identity1}) give 
\begin{equation}
\int_{-\infty}^{\infty}e^{-(\beta\gamma+\hbar)
[(\beta\gamma-\hbar)x-(\beta\gamma+\hbar)y]^2/(4\beta^2\hbar^2)} \,H_n(
y\sqrt{\gamma/\beta\hbar}) \,dy = \beta\sqrt{\pi}\,
\lambda_n^{(G)}(\beta,\gamma) \,H_n( x\sqrt{\gamma/\beta\hbar})\,.
\label{identity2}
\end{equation}
This is a known identity \cite{Gradshteyn00} and provides a check on the
validity of (\ref{gd09}). 

The spectral bounds on $\hat{\rho}_{\beta,\gamma}^{(G)}$, are graphed in
fig.~\ref{fig01} against the classical uncertainty $\Delta q\Delta p$ (in units
of $\hbar$). When $\Delta q \Delta p < \hbar/2$, we obtain the strongly
classical Groenewold operators discussed above. At the point $\Delta q \Delta p
= \hbar/2$, the Groenewold operator has only one non-zero eigenvalue
$\lambda_0=1$, and is precisely the pure state density $|0\ra\la 0|$. For
$\Delta q \Delta p$ above this critical value, the spectrum is uniformly
positive and bounded by 0 below and by $\lambda_0 = 2\hbar/(2\hbar +
\beta\gamma) <1$ above. As such, for $\beta\gamma > \hbar$, the Groenewold
operator $\hat{\rho}_{\beta,\gamma}^{(G)}$ may be considered as the mixed state
quantum density
\begin{equation}
\hat{\rho}_{\beta,\gamma}^{(G)} \,=\, \sum_{n=0}^\infty p_n |n\ra\la n|\,,\quad p_n
\equiv \lambda_n^{(G)}(\beta,\gamma) >0\,,\quad \sum_{n=0}^\infty p_n =1\,.
\end{equation}

\end{section} 
 
\begin{section}{Uniform Densities} 
 
In the Gaussian Case (A) above, there seems to be an  obvious connection
between the  value of the uncertainty product associated with the  Liouville
density and the appearance of negative  eigenvalues of the corresponding
Groenewold operator. This  is not necessarily the case for more general
densities, however, and we illustrate  this point by considering Case (B), the
family of uniformly distributed Liouville densities, defined on the interior
of an ellipse in $\Gamma$, centred at the origin: 
\begin{equation} 
\rho_{\beta,\gamma}^{(U)}(q,p) \,=\, \left\{ \begin{array}{ccc} \frac{1}{\pi \beta \gamma } \,,
&& 0\leq \tfrac{q^2}{\beta^2} + \tfrac{p^2}{\gamma^2} \leq 1\,, \\ \\
0& & \tO\,. \end{array} \right. 
\end{equation} 
The uncertainty product associated with $\rho_{\beta,\gamma}^{(U)}(q,p)$ 
is $\Delta q\Delta p=\beta\gamma/4$.  
 
The corresponding Groenewold operators are given by 
\begin{equation} 
\hat{\rho}_{\beta,\gamma}^{(U)} \,=\, \int_\Gamma
\rho_{\beta,\gamma}^{(U)}(q,p) \hD(q,p)  \,dq \,dp\,. 
\end{equation} 
We recognise these as scaled versions of operators that have been considered
previously in a different context \cite{Bracken99}. They also commute with
$\hat{N}$ as defined above and their eigenvalues may be expressed as
\begin{equation} 
\lambda_{n}^{(U)}(\beta, \gamma) \,=\, \frac{2(-1)^n}{\beta\gamma}
\int_0^{\beta\gamma}  e^{-t} L_n(2t)\, dt\,,\quad n=0,\,1,\,2,\,\dots\,. 
\end{equation} 
In fig.~\ref{fig01}, the bounds on the spectrum of
$\hat{\rho}_{\beta,\gamma}^{(U)}$ are  graphed against $\beta\gamma$ (in units
of $\hbar$), from which one sees that although the bounds decrease in magnitude
as $\Delta q \Delta p$ rises above  $\hbar/2$, the lower bound remains below
zero. 
 
There is a simple physical explanation for this phenomenon: by constructing a 
uniform distribution over a finite subregion of $\Gamma$, one is restricting 
both the position and momentum of a particle to a finite interval. This is
impossible for a quantum system because the configuration and momentum space
states are related by the Fourier transform and it is known that the Fourier
transform of a function with compact support is entire \cite{Titchmarsh48}.
This argument applies not only to Case (B) but to any Liouville density with 
compact support in $\Gamma$: for any such density, the corresponding Groenewold
operator cannot be a true quantum density and it follows that since it has a
trace equal to 1, it must have at least one negative eigenvalue in its
spectrum. However, we note that for the example considered here, the negative
values decrease in amplitude as the area of the ellipse increases.

\begin{figure} 
\centerline{\psfig{figure=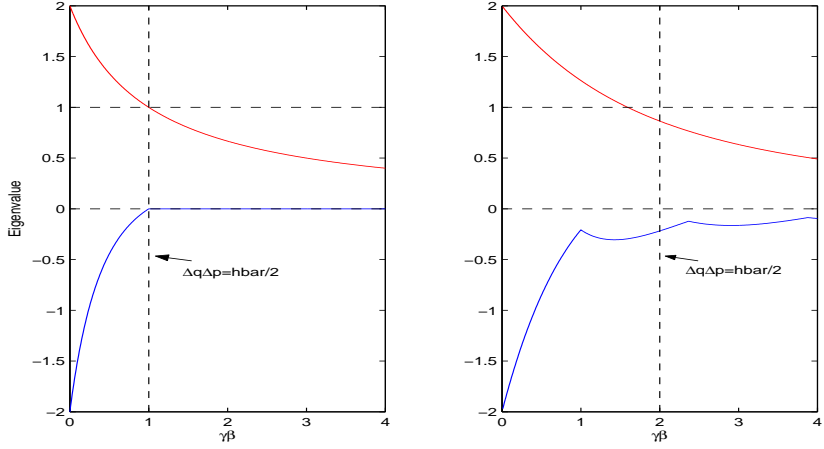,height=60mm,width=110mm}} 
\caption{Spectral bounds on $\hat{\rho}_{\beta,\gamma}^{(G)}$ (left) and
$\hat{\rho}_{\beta,\gamma}^{(U)}$ (right). Note that when the classical
uncertainty product exceeds $\hbar/2$, the Groenewold operators for the
Gaussians considered in Case (A) are non-negative definite. On the right we
show that no such transition occurs for the Groenewold operators corresponding
to the uniform distributions considered in Case (B).} 
\label{fig01} 
\end{figure}

\end{section} 
 
\begin{section}{Conclusion} 

Just as Wigner functions play a central role in the phase space formulation of
quantum mechanics, so Groenewold operators play a central role in the Hilbert
space formulation of classical mechanics,  which offers the possibility of new
insights into the classical-quantum interface.  It is therefore important to
elucidate the properties of Groenewold operators. In the present work, we have
considered exactly solvable examples to illustrate some specific properties. We
have seen that  ``violation of the uncertainty principle" and  restriction to a
compact subregion of $\Gamma$ can lead to eigenvalues outside of the range
$[0,1]$, but this can occur due to the influence of other factors. The
emergence of negative eigenvalues when Groenewold operators that are initially
true density operators are allowed to evolve classically rather than quantum
mechanically \cite{Muga92,Habib02} is a prime example. There is evidence
\cite{Habib02} to suggest that when the effects of the  classical and quantum
evolution diverge rapidly (such as for a  delta-kicked rotator), the magnitude
of the negative eigenvalues is  much greater than for slowly diverging
evolutions (such as for the Duffing oscillator). In recent work, we have shown
that large negative eigenvalues arise in the classical evolution of an initial
coherent state density \cite{Bracken04} in an anharmonic potential
\cite{Milburn86} that causes the phase space density to develop ``whorls". 

\end{section}

\begin{section}{Acknowledgments}
The authors thank C. Zachos for helpful comments.
\end{section}

\end{document}